\documentclass[12pt]{article}
\usepackage{amssymb}
\usepackage{amsmath}
\usepackage{amsfonts}
\usepackage{epsfig}
\usepackage{anysize}
\marginsize{2cm}{2cm}{2.5cm}{3cm}
\date{}

\begin{document}

\title{\LARGE \bf Relating high dimensional stochastic complex systems to low-dimensional intermittency}
%\shorttitle{Distributions of sums of chaotic positions.}

%\author{Alvaro Diaz-Ruelas\inst{1,2}\thanks{alvaropdr@fisica.unam.mx} \and Henrik Jeldtoft Jensen\inst{3}\thanks{h.jensen@imperial.ac.uk} \and Duccio Piovani\inst{3,4}\thanks{duccio.piovani@gmail.com} \and Alberto Robledo\inst{1,2}\thanks{robledo@fisica.unam.mx}}  
%
%\affiliation{Instituto de F\'isica, Universidad Nacional Aut\'onoma de M\'exico, Ciudad Universitaria, Ciudad de M\'exico, 04510, Mexico \and Centro de Ciencias de la Complejidad, Universidad Nacional Aut\'onoma de M\'exico, Ciudad Universitaria, Ciudad de M\'exico, 04510, Mexico \and Centre for Complexity Science and Department of Mathematics, Imperial College London, South Kensington Campus, SW7 2AZ, UK \and Centre for Advanced Spatial Analysis. University College London, W1T 4TJ, London, UK} 

\author{Alvaro Diaz-Ruelas\textsuperscript{1,2}\thanks{alvarodiaz@ciencias.unam.mx}, Henrik Jeldtoft Jensen\textsuperscript{3}\thanks{h.jensen@imperial.ac.uk},  Duccio Piovani \textsuperscript{3,4}\thanks{duccio.piovani@gmail.com}, Alberto Robledo\textsuperscript{1,2}\thanks{robledo@fisica.unam.mx}\vspace*{12pt}\\ 
\footnotesize 1. Instituto de F\'isica, Universidad Nacional Aut\'onoma de M\'exico, \\ \footnotesize Ciudad Universitaria, Ciudad de M\'exico, 04510, Mexico  \\ \footnotesize 2. Centro de Ciencias de la Complejidad, Universidad Nacional Aut\'onoma de M\'exico, \\ \footnotesize Ciudad Universitaria, Ciudad de M\'exico, 04510, Mexico \\ \footnotesize 3. Centre for Complexity Science and Department of Mathematics, Imperial College London, \\ \footnotesize South Kensington Campus, SW7 2AZ, UK \\ \footnotesize 4. Centre for Advanced Spatial Analysis. University College London, W1T 4TJ, London, UK}

\maketitle

%%%%%%
\abstract{
We evaluate the implication and outlook of an unanticipated simplification in the macroscopic behavior of two high-dimensional sto-chastic models: the Replicator Model with Mutations and the Tangled Nature Model (TaNa) of evolutionary ecology. This simplification consists of the apparent display of low-dimensional dynamics in the non-stationary intermittent time evolution of the model on a coarse-grained scale. Evolution on this time scale spans generations of individuals, rather than single reproduction, death or mutation events. While a local one-dimensional map close to a tangent bifurcation can be derived from a mean-field version of the TaNa model, a nonlinear dynamical model consisting of successive tangent bifurcations generates time evolution patterns resembling those of the full TaNa model. To advance the interpretation of this finding, here we consider parallel results on a game-theoretic version of the TaNa model that in discrete time yields a coupled map lattice. This in turn is represented, a la Langevin, by a one-dimensional nonlinear map. Among various kinds of behaviours we obtain intermittent evolution associated with tangent bifurcations. We discuss our results.}

%\pacs{05.45.Ac, 87.23.Cc, 87.23.Kg}
%05.45.Ac Low dimensional chaos
%05.45.-a Nonlinear dynamics and chaos
%87.23.Cc Population dynamics and ecological pattern...
%87.23.Kg Dynamics of evolution

%\keywords{}

%%%%%%%%%%%%%
%%INTRODUCTION%%
%%%%%%%%%%%%%

\section{Introduction} 
A characteristic feature of a complex system constituted by many individual agents or degrees of freedom is the occurrence of different levels of behavior separated by differing spatial and time scales. The connections between these levels are intriguingly complex due to the presence of nonlinearities. But this very presence makes difficult, if not impossible, the decomposition of the dynamics of the high-dimensional problem into many independent ones, like the so-called normal modes of linearized dynamics. An alternative approach to the investigation of complex systems is to capture their important properties by means of simplified high-dimensional models (\textit{i.e.} those involving many interacting individual agents), as it is \textit{e.g.} done in the Tangled Nature (TaNa) model of evolutionary ecology \cite{tana:article1}. This is clearly also the spirit that led to the definition of coupled map lattices (CMLs) introduced independently in the 1980s by Kaneko \cite{Kaneko_1984} and Kapral \cite{Kapral_1986}. A CML is a system with discrete space and time variables but with a continuous local variable described by a nonlinear function. Therefore it is a collection of a large number of coupled nonlinear low-dimensional iterated maps.

Interestingly, macroscopic collective behavior can arise in CMLs as shown convincingly some time ago by Chat{\'e} and Manneville \cite{Chate_Manneville_1992}. Namely, the emergence of low-dimensional behavior in the coarse-grained description of the high-dimensional dynamics. But very recently another example of this occurrence has been exhibited for the high-dimensional TaNa model under mean-field approximations \cite{ADR_HJJ_DP_AR1}, such that effective low-dimensional dynamics is displayed in the macroscopic non-stationary intermittent evolution of the model. Recently, a game-theoretic rendering of the TaNa model described by a set of coupled replicator equations that incorporate stochastic mutations has been derived and studied \cite{DP_JG_HJJ_2016}, and found to exhibit macroscopic non-stationary intermittent evolution similar to that in the TaNa model. Noticeably, in relation to the above, the discrete time version of the game-theoretic model that operates in the limit of many strategies constitutes a coupled map lattice.   

The next step we consider is a radical simplification of the game-theoretic CML replicator-mutation equations into a one-dimensional nonlinear map. This attempt aims at probing the possible connection between the macroscopic intermittent behaviors of the above-mentioned high-dimensional models with the low-dimensional established sources of intermittency, such as the tangent bifurcation \cite{schuster1} with known $1/f$ noise spectra \cite{schuster2}. In doing so, we reduce the many-strategy game-theoretic problem to a classic version of two strategies, where one of them represents the chosen agent or species and the other assemblages all the others. Finally, we recall that a one-dimensional nonlinear dynamical model can be constructed \cite{ADR_HJJ_DP_AR1} such that its time evolution consists of successive tangent bifurcations that generate patterns resembling those of the full TaNa model in macroscopic scales. The parameters in the model are based on identified mechanisms that control the duration of the basic quasi-stable event generated by a local mean-field map derived from the TaNa model \cite{ADR_HJJ_DP_AR1}.

%{\color{blue} The justification of this procedure is akin to the Langevin theory \cite{Chaikin_Lubensky_1995}, that finds a way round the detailed consideration of many degrees of freedom by representing via a single term, say, a noise source, the effect of their interactions on an individual one. In the same spirit, attractors of nonlinear low-dimensional maps under the effect of external noise can be used to model behaviors in systems with many degrees of freedom such as CMLs \cite{baldovin1}}.

%{\color{blue}In the body of the paper we first describe the CML replicator model with mutations, its approximation to a two-strategy replicator-mutation equation that constitutes a one-dimensional nonlinear map \cite{DV_AR_AS_2011}, and present the occurrence of tangent bifurcations that links with intermittent behavior. In the second part we recall how the TaNa model under mean field and local approximations leads to a one-dimensional map near a tangent bifurcation and then refer to the consecutive tangent bifurcation model and the succeeding intermittent patterns that it typically generates. We finish with a brief summary and discussion that expresses our motivation to encourage a redevelopment of a closer relation between nonlinear dynamical systems research and the science of complex systems.}     

\section{The Replicator Model with Mutations}
Here we briefly present the intermittent behavior of the Replicator Model with Mutations, details of which can be found in \cite{DP_JG_HJJ_2016}. The replicator equation~\cite{taylor1978evolutionary} was introduced in evolutionary game theory in order to capture the frequency-dependent nature of the evolution process. We are interested in the limit of many strategies. Players may leave the system (say go bankrupt or extinct) or may change their strategy (mutate). This means that the number of players choosing a given strategy and the number of available strategies are in constant evolution. This version of the replicator dynamics set-up was studied by Tokita and Yasutomi in~\cite{tokita2003emergence}. The authors focused on the emerging network properties. Here we continue this study but with an emphasis on the intermittent nature of the macro-dynamics.

The model is described in terms of a configuration vector $\bf{n}(t)$ which contains the \emph{relative frequencies} of all the allowed $d$ different strategies, so the components $n_i(t)\in[0,1]$ for all $i=1,2,...,d$. A $d\times d$ payoff matrix $J$ contains the payoffs of every pairwise combination. The matrix $J$ is a random and fixed interaction network on top of which the replicator dynamics will evolve. Each strategy distinguishes itself from the others in its payoffs or interactions with the rest of the strategy space. We used an uncorrelated, large matrix of dimension  $d\in ( 10^2 ,10^4)$. In the initial configuration, $N_{o}<d$ strategies start with the same frequency $n_i =1/N_{o}$ and for all the other strategies $n_i(0)=0$. The empty strategies can become populated only by one of the \emph{active} strategies mutating into them. Once this happens, their frequency will evolve according to the replicator equation, Eq. (\ref{Eq:ReplicatorEq}), in which these newly occupied strategies interact with the active strategies which they are linkedf through the matrix $J$. 

A time step of the replicator dynamics consists of calculating the \emph{fitness}, $h_i(t) =  \sum_j J_{ij}n_j(t) $ of each active strategy and compare it with the average fitness $\bar{h}(t)= \sum_{ij}J_{ij}n_{i}(t)n_{j}(t) $. The frequencies are  updated according to
\begin{equation}
n_{i}(t+1)= n_i(t)+\left( \sum_j J_{ij} n_j(t) - \sum_{k,j}J_{kj}n_{k}(t)n_{j}(t) \right) n_i(t).
\label{Eq:ReplicatorEq}
\end{equation}
When the frequency of a strategy $i$ goes below a preset extinction threshold $n_i(t)<n^{{ext}}$, the strategy is considered extinct and its frequency is set to zero $n_i(t+1) = 0$. Right after an extinction event the system is immediately renormalised in order to maintain the condition $\sum_i n_i(t) =1$.

The stochastic element consists in the following updates. With probability $p^{{mut}}$ each strategy mutates into another one, this is done by transferring a fraction $\alpha_{{mut}}$ of the frequency from the considered strategy to another strategy. The label of the latter strategy is chosen in the vicinity of the first by use of a normal distribution $N(i,\Delta)$ centred on label $i\in\{1,2,...,d\}$ with variance $\Delta$ with periodic boundary conditions, i.e. label $d+1$ is identified with label $1$. The closer the labels of two strategies are the more likely it is for one to mutate into the other.

\begin{figure}[!h]
\centering
%%\captionsetup{width=\linewidth}
\includegraphics[width=0.47\columnwidth]{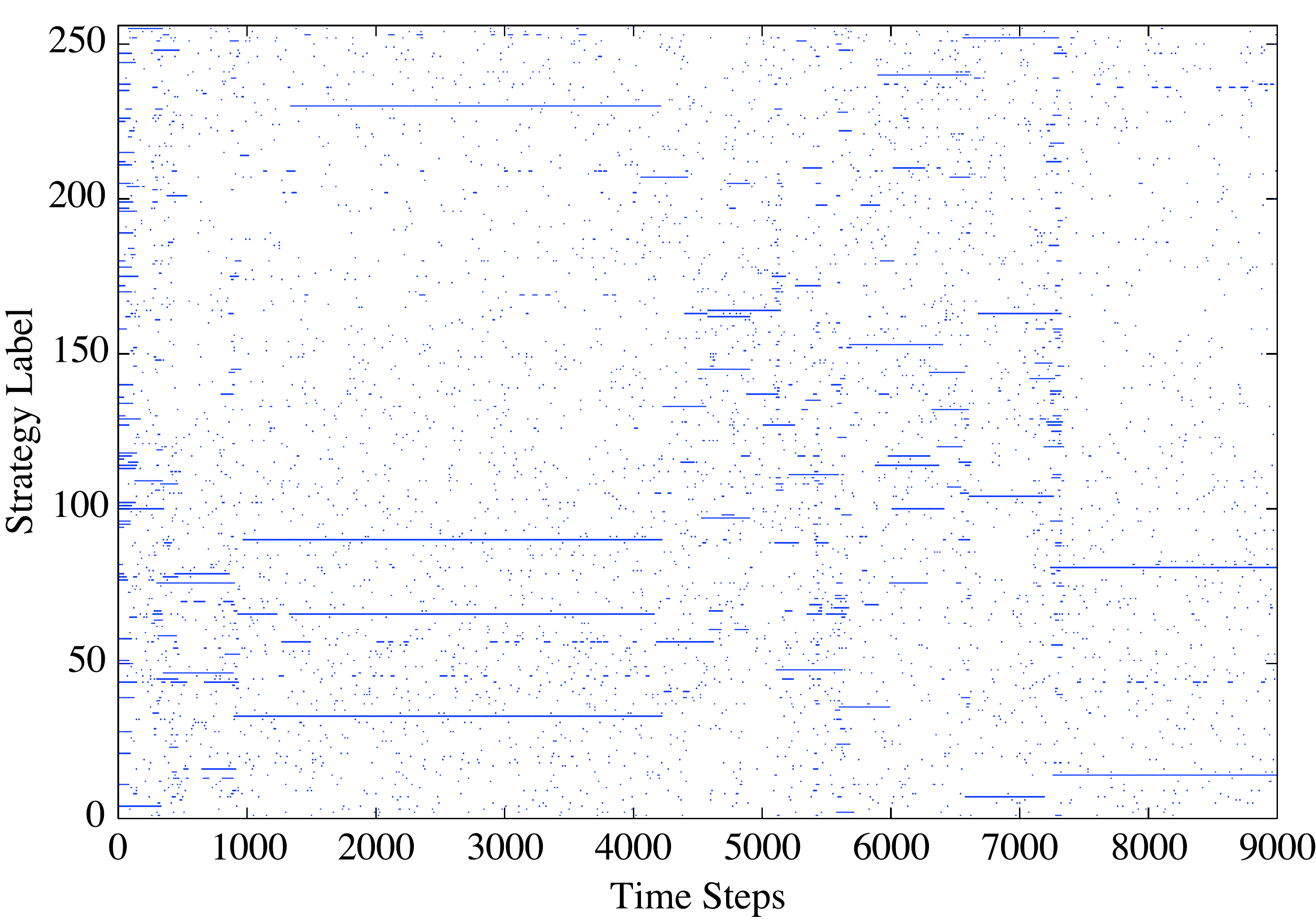}
\includegraphics[width=0.47\columnwidth]{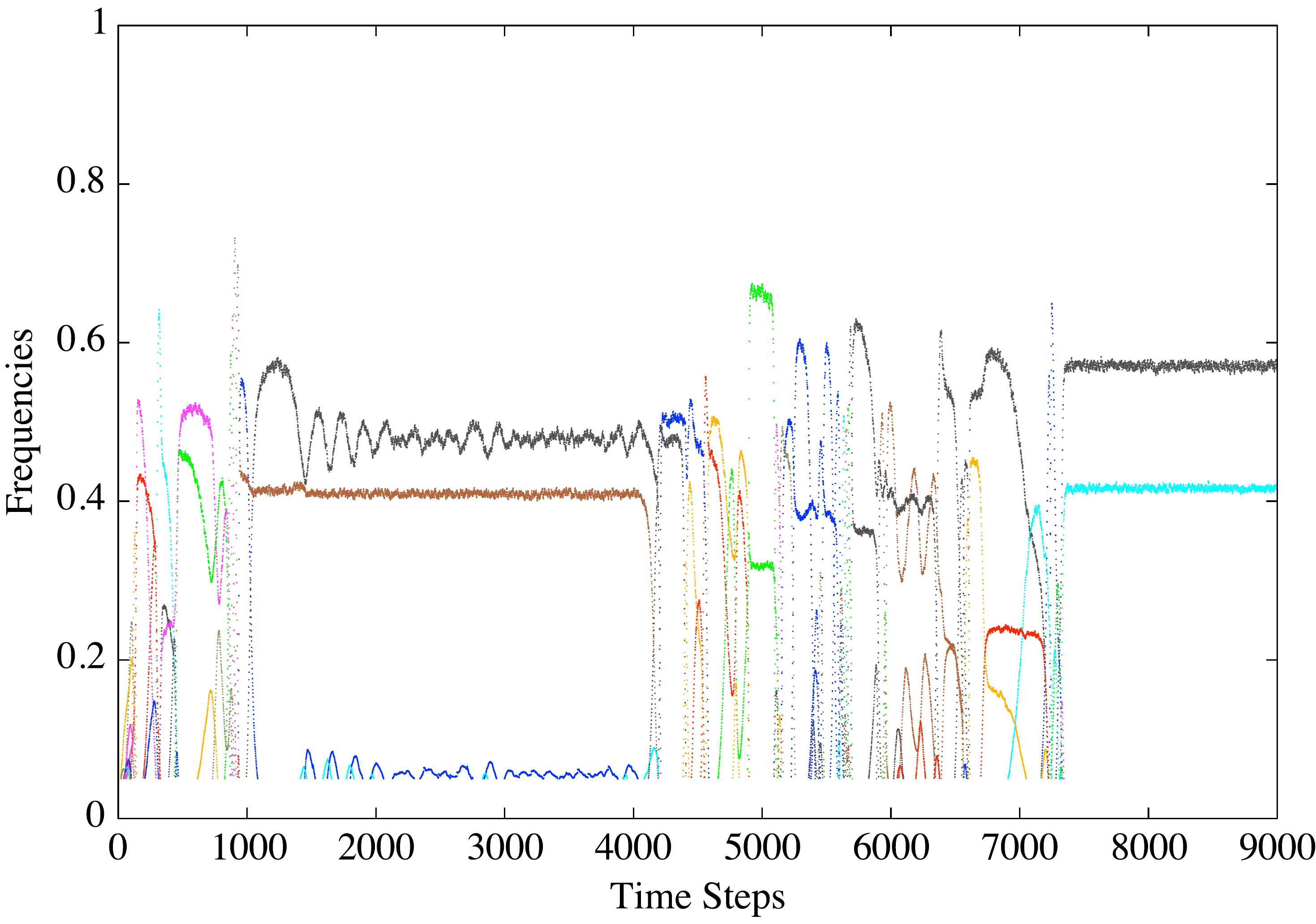}
\caption{Left panel: occupancy distribution of the types. The genotypes are labelled arbitrarily and a dot indicates a type which is occupied at the time $t$ (\textit{i.e.} $\textbf{n}(t)>0$). The punctuated dynamics is clearly visible: quasi-stable periods alternate with brief periods of hectic transitions. Right panel: the frequencies of the strategies. Each colour belongs to a different strategy. Once again the transitions from one meta stable configuration (approximate fixed point) to another is clear.}
\label{fig:2}
\end{figure}

The systemic level dynamics is described by $\textbf{n}(t)$ and is shown in fig.(\ref{fig:2}), where we present the \emph{occupancy} plot (left panel) and the evolution of the frequencies of the single strategies (right panel).

The parameters used in the simulations are $d = 256$, $n_{ext}=0.001$, $\alpha_{mut}=0.01$, $p^{mut} =0.2$ and $\Delta=15$  and were chosen for reasons of computational performance. The meta stable states are typically characterised by two strongly occupied strategies which are surrounded by 7 to 8 ``cloud" strategies. These are populated by mutations and quickly die out.  

\subsection{Mean Field Description}

The random mutations are the only source of stochasticity in the model's dynamics. These stochastic events can make the frequency of a strategy grow as a result of inflow from different strategies mutating on to the given strategy and can lead to     a strategy looses part of its frequency due to mutations onto other strategies. 
    The gain is on average given by $\alpha_{{mut}} n_j(t+1)$ which happens with probability $p_{mut}\sum_{j\in N_a} p_{j\rightarrow i}$, where $N_a$ is the number of active strategies and
\begin{equation}p_{j \rightarrow i} = \frac{e^{\frac{-|i-j|^2}{2\Delta^2}}}{\sqrt{2\pi \Delta^2}}
\end{equation}
is the probability of $i$ mutating into $j$ (and viceversa). A fraction of players $\alpha_{mut}$ are lost, which happens with probability $p_{mut}$. We therefore get the mean field description as 
\begin{eqnarray}
n_{i}(t+1)\simeq n_i(t)+\left( \sum_j J_{ij}n_j(t) - \sum_{jk}J_{ik}n_{i}(t)n_{k}(t) \right)n_i(t)\cr
 + p_{mut} \alpha_{{mut}}\left( \sum_j n_j(t) p_{j \rightarrow i } -  n_i(t) \right).
\label{eq:mf}
\end{eqnarray}

\begin{figure}[!h]
\centering
\includegraphics[width=0.95\columnwidth]{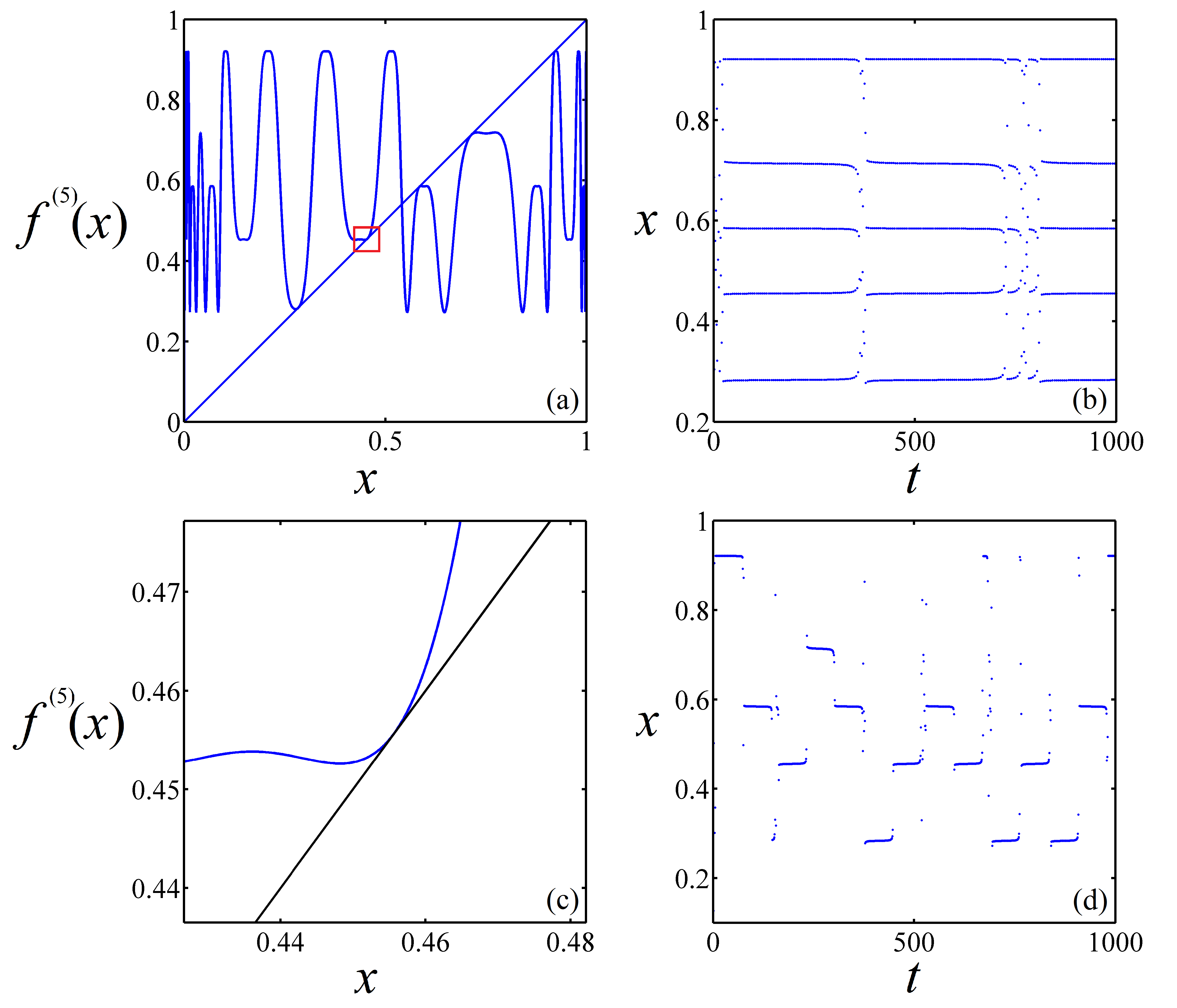}
\caption{Simulation of the replicator map in Eq. (\ref{eq:replicator_map}) near a period five periodic window with control parameters $S=T=6.5950$. Panel (a): The map (Eq. (\ref{eq:replicator_map})) composed five times. Panel (b): Trajectory obtained from the map without being composed, $f(x)$, showing the laminar episodes separated by chaotic bursts. Panel (c): Enlargement of the map in (a) showing a local near-tangent piece. See the enclosed region inside the square in Panel (a). Panel (d): Trajectory obtained form the map $f^{(5)}(x)$ showing the laminar episodes separated by chaotic bursts. The value of the Lyapunov exponent is $\lambda=0.001391$.}
\label{rep_map}
\end{figure}

We now bravely reduce Eq. (\ref{eq:mf}) to a one dimensional map intending to capture the evolution of the occupancy of a single strategy as it evolves and interact with all other strategies and arrive at
\begin{equation}
	n(t+1) = n(t)+J_1n^2(t)+J_2n^3(t)+\alpha n(t).
\end{equation}
Here $J_1$ represent the average effect of the $J_{ij}n_j(t)$ term in Eq (\ref{eq:mf}), $J_2$ the effect of the $J_{ik}n_{i}(t)n_{k}(t) $ term and $\alpha$ sums up the effect of the last term in the equation.  This  mean field equation is of the same form as 
\begin{equation}
	x_{k+1}=f(x_k)=x_k+\delta x_k(1-x_k)[S+(1-T-S)x_k], \label{eq:replicator_map}
\end{equation}
with $\alpha=\delta S$, $J_1=\delta(1-T-2S)$ and $J_2=\delta(1-T-ST)$. This map has been studied in detail in \cite{DV_AR_AS_2011}. Note we have included a factor $\delta$ (omitted in \cite{DV_AR_AS_2011}) to represent the size of the time step when going from Eq. (1) to Eq. (2) in Ref. \cite{DV_AR_AS_2011}. Here we simply present simulations in Fig. \ref{rep_map} to demonstrate that the map can reproduce behavior very similar to the simulation of the full model. 

The fact that the intermittency of the high dimensional Replicator Model with Mutations may be qualitatively related to tangent bifurcation of a one dimensional map encourage us to discuss in the next section a similar strategy of dramatic dimensional reduction for the fully stochastic Tangled Nature model.  

\section{The Tangled Nature model}
The Tangled Nature model is a model of evolutionary ecology, which studies the macro-dynamics emerging from the dynamics of individual agents, co-evolving in a web of mutual interactions. The systemic level dynamics exhibit intermittency. The model was introduced in \cite{tana:article1,tana:article2} and since then, the model framework has been used by several authors see \textit{e.g.} \cite{FakeTanaBasic,FakeTaNaFluctuations,Becker_Sibani_2014,Nicholson_Sibani_2015,Vazquez_2015}. A summary of some of the models features and predictions can be found in \cite{Sibani2013}.    

\subsection*{Description of the model}
In contrast to the replicator model studied above the Tangled Nature model is fully stochastic. Here is briefly how the model is updated. The dynamical entities of the TaNa model consist of  agents represented by a sequence of binary variables with fixed length $L$ \cite{Higgs/Derrida:article}. We denote by $n({\bf S}^a,t)$ the number of agents of type ${\bf S}^a=(S^a_1,S^a_2,...,S^a_L)$ (here $S^a_i\in\{-1,1\}$) at time $t$ and the total population is $N(t)=\sum_{a=1}^{2^L} n({\bf S}^{a},t)$. 
A time step is defined as a succession of one annihilation and of one reproduction attempt. Annihilation consists of choosing an agent at random with uniform probability and then removing the agent with probability $p_{kill}$, taken to be constant in time and independent on the type. Reproduction: choose with uniform probability an agent, $\textbf{S}^a$,  at random and duplicate the agent (and remove the mother) with probability
\begin{equation}
p_{off}({\bf S}^a,t)=\frac{\exp{(H({\bf S}^a,t))}}{1+\exp{(H({\bf S}^a,t))}},
\label{eq:2.3}  
\end{equation}
which depends on the occupancy distribution of all the types at time $t$ through the weight function
 \begin{equation}
H({\bf S}^a,t)= \frac{k}{N(t)}\sum_b J({\bf S}^a,{\bf S}^b)n_b(t) -\mu N(t).
\label{eq:1}
\end{equation}
In Eq. (\ref{eq:1}), the first term couples the agent ${\bf S}^a$ to one of type ${\bf S}^b$ by introducing the interaction strength $J(\mathbf{S}^a,\mathbf{S}^b)$, whose values are randomly distributed in the interval $\left[-1,+1\right]$. For simplification and to emphasize interactions we here assume: $J(\mathbf{S}^a,\mathbf{S}^a)=0$. The parameter $k$ scales  the interactions strength and $\mu$ can be thought of as the carrying capacity of the environment. An increase (decrease) in $\mu$ corresponds to harsher (more favourable) external conditions.

\begin{figure}[ht!]
\centering
\includegraphics[width=0.45\columnwidth]{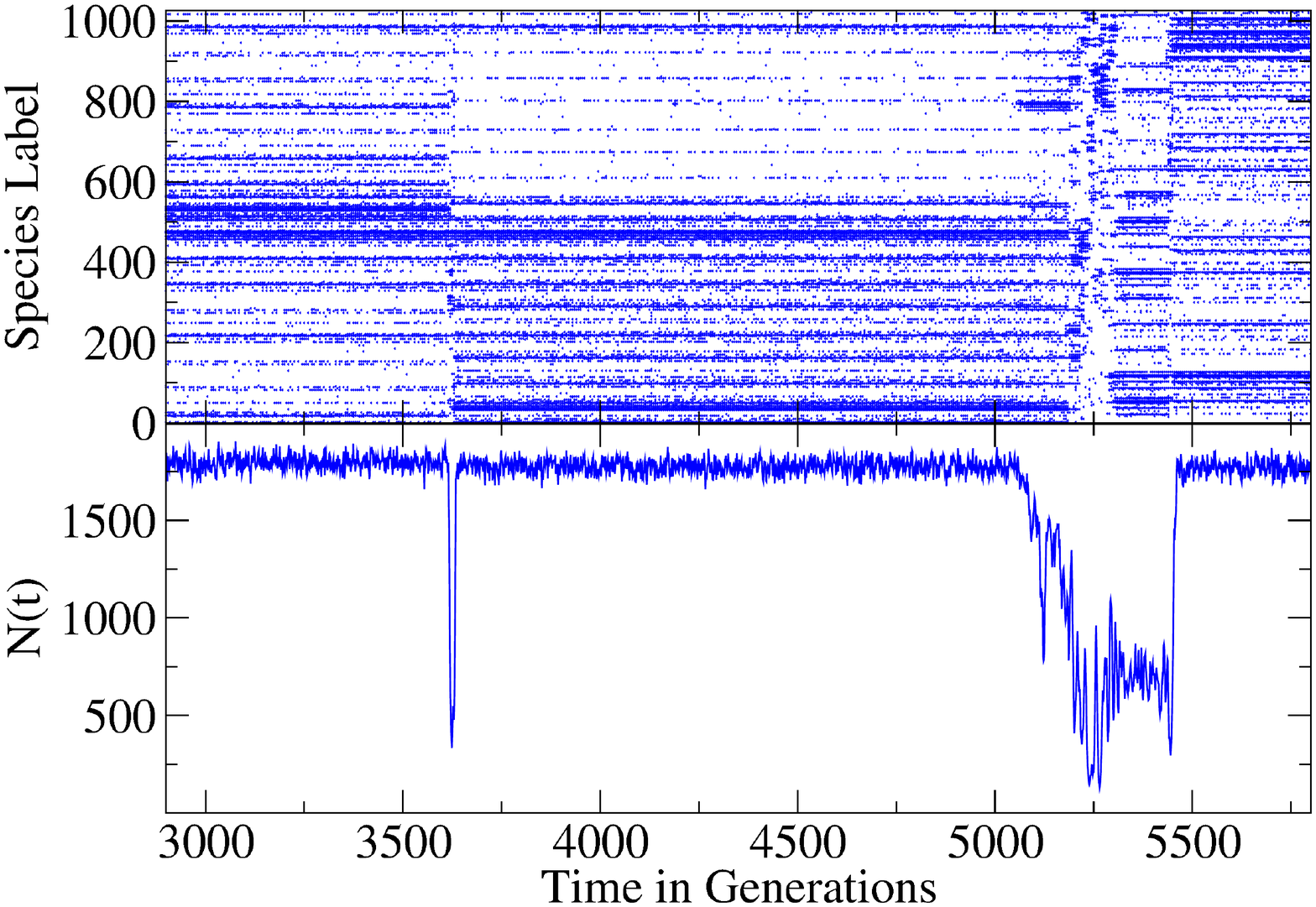}%[width=4.5cm,height=6.67cm]{N_O.pdf}
\includegraphics[width=0.45\columnwidth]{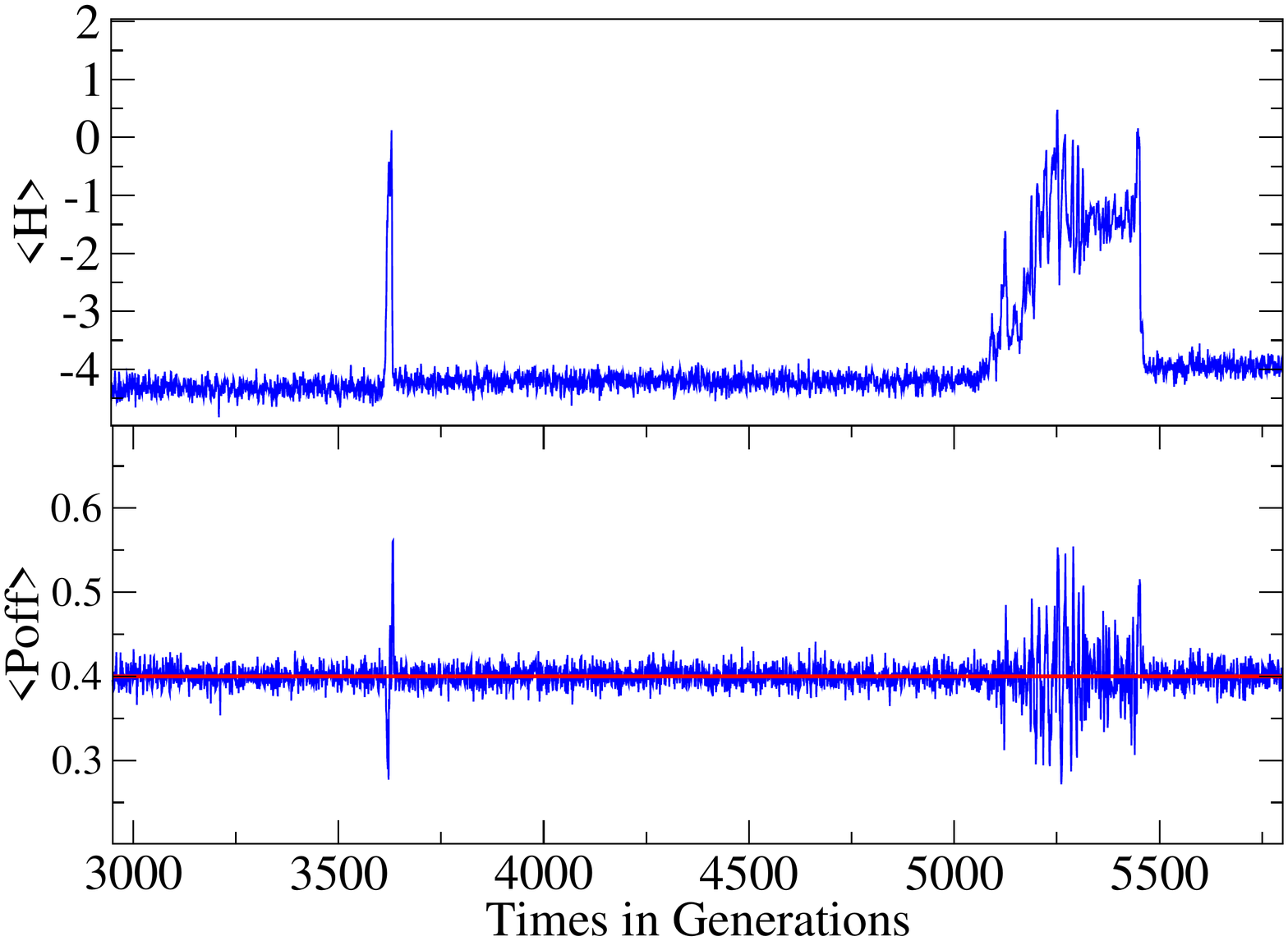}%[width=4.5cm,height=6.67cm]{H_P.pdf}
\caption{Left Panel: Total population as a function of time (in generations) for a single realization of the TaNa model. The punctuated dynamics is clearly visible: quasi-stable periods alternate with periods of hectic transitions, during which $N(t)$ exhibits large amplitude fluctuations. Right panel: The averages of the weight function $H$ and the reproduction probability $p_off$. The parameters are $L = 10$, $p_{kill}  = 0.4$, $p_{mut} = 0.02$, $\mu  = 0.007$, $k = 40$ the red line indicates $p_{kill}$.}
\label{fig:1}
\end{figure}

Mutations occur in the following way: For each of the two copies ${\bf S}^{a_1}$ and ${\bf S}^{a_2}$, a single mutation changes the sign of one of the genes: $ S^{a_1}_i\rightarrow -S^{a_1}_{i}$, $ S^{a_2}_i\rightarrow -S^{a_2}_{i}$ with probability $p_{mut}$. We define a generation to consist of $N(t)/p_{kill}$ time steps, \textit{i.e.} the average time needed to kill all the individuals at time $t$. These microscopic rules generate intermittent macro dynamics\cite{tana:article2} as shown in Fig. \ref{fig:1}. The long quiescent epochs are called quasi Evolutionary Stable Strategies (qESS), since they do remind one of John Maynard Smith's notion of Evolutionary Stable Strategies  introduced in his game theoretic description of evolution \cite{Maynard:book}. 

The weight function $H$ will fluctuate about the value given by the stable dynamical fixed point condition $p_\text{off}(H)=p_{kill}$. This suggests that the mean field value of $H$ may indeed evolve in an intermittent way that may be captured by a tangent map. We will therefore derive the mean field map for $\langle H\rangle$. 
  
\subsection*{Derivation of Mean Field map for $H$}  
Since the model is fully stochastic, the derivation of the mean field equations for the model has to estimate the on-average effect of reproduction, with or without mutations and stochastic death processes. A detailed account of this derivation is given in \cite{ADR_HJJ_DP_AR1}. The final mean field map for the average $\langle H\rangle$ is of the form 
\begin{equation}
	\langle H\rangle \mapsto  \langle H\rangle +A\langle p_{off}\rangle_{ext}	
	-   B p_{kill},
	\label{TaNa_mf_map1}
\end{equation}
	where the coefficients are given by	
\begin{eqnarray}
	A &=& \left(\frac{k\bar{J}}{N}-\mu\right)(1-p_{mut})^{2L}+\left(\frac{k\tilde{J}}{N}-\mu\right)(P^{(0)}_{mut}+Lp^{(1)}_{mut})Lp^{(1)}_{mut}
	\label{A_cof}\\
	B &=& \frac{k\bar{J}}{N}-\mu,
\end{eqnarray}
and depends on the probability for no mutation $P^{(0)}_{mut}$, one mutation $p^{(1)}_{mut}$ and the $J$ coupling averaged over the highly occupied types of agents ${\bar J}$ and the average of the the couplings over these types and the set of types they are connected to via single mutations $\tilde{J}$. The details are in \cite{ADR_HJJ_DP_AR1}.

To obtain a closed expression we expand $p_\text{off}(H^i)$ in Eq. (\ref{TaNa_mf_map1}) to second order about $x^*$ and replace only $\langle H^2\rangle_{ext}$ by $\langle H\rangle^2_{ext}$. This leads to a tangent map. To study the intermittency of this map expand $p_\text{off}(H)$ in Eq. (\ref{TaNa_mf_map1}) to second order about $H^*=\ln [p_{kill}/(1-p_{kill}) ]$,
\begin{equation}
	p_\text{off}(H) = a_0+a_1(H-H^*)+a_2(H-H^*)^2
	\label{2nd_ord}
\end{equation}
where 
\begin{eqnarray*}
	a_0&=&p_{kill},\\
	a_1&=&p'_\text{off}(H^*)=p_{kill}(1-p_{kill}),\\
	a_2&=&\frac{1}{2}p''_\text{off}(H^*)= \frac{1}{2}a_1(1-2p_{kill}).
\end{eqnarray*}
The characteristic time to pass through the narrow passage where the map is parallel to the tangent straight line is to lowest order in the killing probability and only including leading order mutation processes well approximated by
\begin{equation}
	\left(\frac{\pi}{T}\right)^2\simeq -\frac{k}{2N}(\bar{J}-\tilde{J})(1-P_0)\left(\frac{k}{N}\tilde{J}-\mu\right)p_{kill}^2.
	\label{T_duration}
\end{equation}

We conclude that our mean field analysis suggests that the length of the qESS, \textit{i.e.} the metastable quiescent epochs, is set by three mechanisms. First the rate of killing. Second the mismatch between the characteristic interaction strength $k\bar{J}$ of the extant types and the carrying capacity as given by the parameter $\mu$ in Eq. (\ref{eq:1}). And thirdly the difference between the typical interaction strength between the extant types, $\bar{J}$ and the typical interaction strength, $\tilde{J}$ across the set of extant and mutant types located in the perimeter of the set of  occupied types. 

It is natural that the duration of the qESS states increases if the rate of killing decreases and it seems also reasonable that the qESS becomes longer if an equilibrium is established between the web of inter-type interactions, as represented by the coupling term in Eq. (\ref{eq:1}), and the demand expressed by the carrying capacity term in the same equation. Finally, loosely speaking at the level of mean field (see Ref. \cite{ADR_HJJ_DP_AR1} for details), if the surrounding mutants originating from the extant types experience interactions significantly different from the existing coupling these mutants may very well be able to out compete existing types and thereby destabilise the current qESS. This is what the term $(\bar{J}-\tilde{J})(1-P^{(0)})$ represents.     

\section{A consecutive tangent bifurcation model}

A simple nonlinear dynamical model is capable of imitating some features of the macroscopic dynamics described above \cite{ADR_HJJ_DP_AR1}. This model makes use of families of chaotic attractors near tangent bifurcations present in low-dimensional iterated maps that display intermittency of type I \cite{schuster1}. These families can be taken from those occurring in quadratic maps, such as the quadratic logistic map,  $f_{\nu }(x)=1-\nu x^{2}$, $-1\leq x\leq 1$, $0\leq \nu \leq 2$. The dynamics at the vicinity $\nu \lesssim \nu _\tau$ of the tangent bifurcation at $\nu =\nu _\tau$ displays intermittency. That is, the map trajectories consist of quasi-periodic motion interrupted by bursts of irregular behaviour. The iteration time duration of the quasi-periodic episodes increases as the tangent bifurcation is approached. At the tangent bifurcation the duration of the episodes diverges and the motion becomes periodic.

A phenomenological procedure for generating successive qESS with durations obtained from the criteria given by Eq. (12) is briefly described as follows \cite{ADR_HJJ_DP_AR1}. First choose a control parameter value $\nu _{0}$ just
left of a window of periodicity $\tau_{0}$ of the logistic map with tangent
bifurcation at $\nu _{\tau_{0}}$, $\delta \nu _{0}\equiv \nu _{0}-\nu
_{\tau_{0}}\lesssim 0$. When the map trajectory with initial condition $x_{0}$
comes out of the bottlenecks formed by $f^{(\tau_{n})}(x)$ and the identity
line to experience a chaotic burst before it is re-injected close to the bottlenecks. The map trajectory evolves in this environment (performing one or more holdup passages and re-injections) until a set of two stochastic conditions is fulfilled, in which case another control parameter value $\nu_1$ is generated just left of a window of periodicity 
$\tau_{1}$ with $\delta \nu _{1}\equiv \nu _{1}-\nu _{\tau_{1}}\lesssim 0$, and so
on for $n=2,3,\ldots $.  These two conditions refer to exceedances associated with two random variables $\delta_1$ and $\delta_2$, distributed by a uniform and a normal distribution, respectively. The conditions are  $\delta_1>\Gamma_1$ and $\delta_2>\Gamma_2$ where  $\Gamma_1$ and $\Gamma_2$ are two prescribed thresholds. Only when the two thresholds are overcome simultaneously the control parameter value is changed to that of a different window, otherwise the trajectory remains close to the same window. The two implemented thresholds correspond to critical values of the imbalances referred after Eq. (\ref{T_duration}),

\begin{equation}
\centering
 \delta_1 = \frac{\frac{k}{N}\tilde{J}-\mu}{p^2_{kill}(1-P_0)} \ \ \mathrm{and} \  \ \delta_2 = \frac{\frac{k}{2N}(\bar{J}-\tilde{J})}{p^2_{kill}(1-P_0)}. 
\end{equation}
\vspace{6pt}

\begin{figure}[ht!]%
\includegraphics[width=0.85\columnwidth]{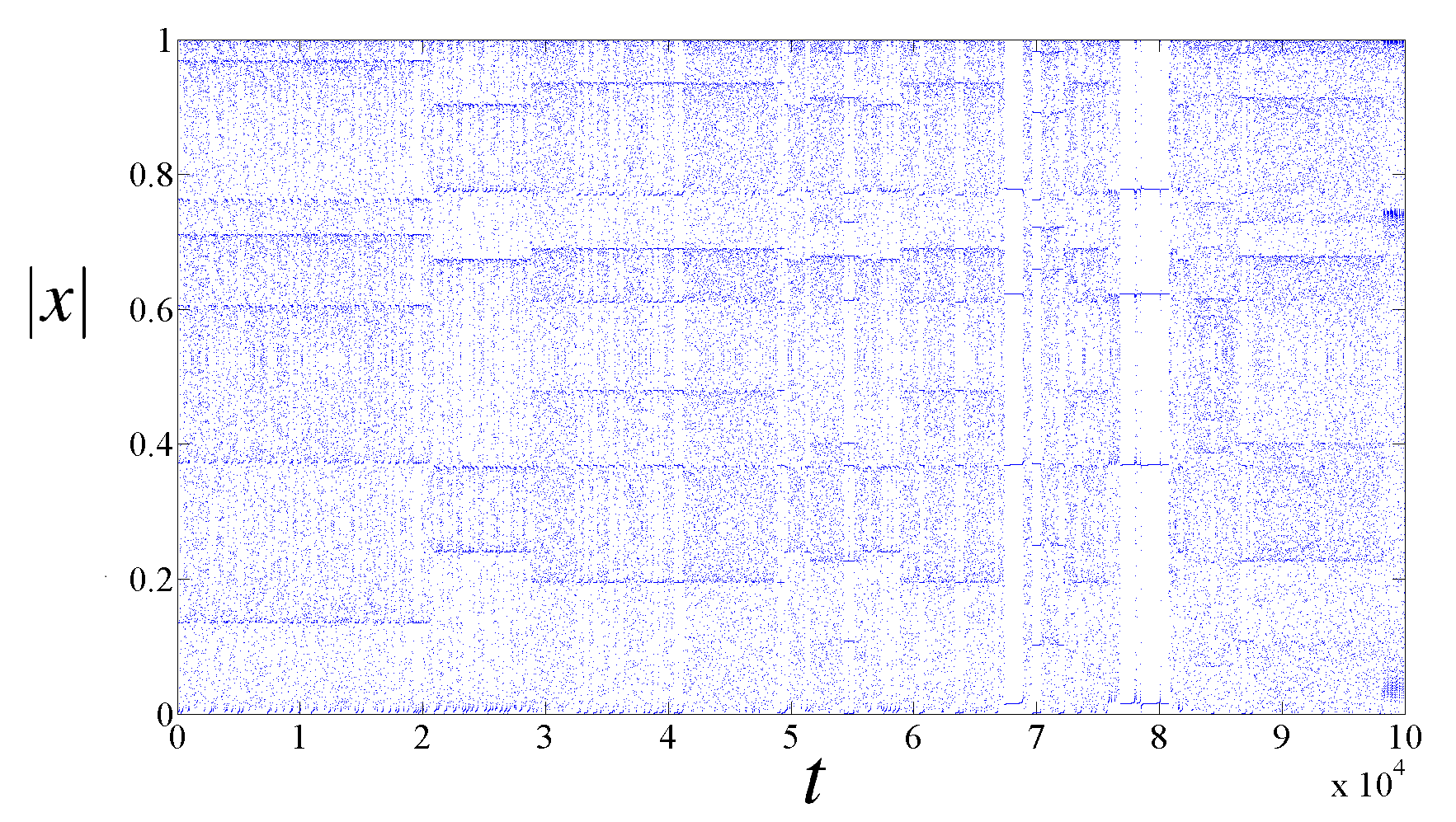}%
\caption{Iterated time evolution of a trajectory generated by the consecutive tangent bifurcation model. The figure is composed of segments, each of which corresponds to a fixed value of the control parameter close to a tangent bifurcation, associated with a given period. Within each segment, many laminar episodes occur separated by chaotic bursts. The periods of the segments are consecutively, from left to right, 8, 7, 9,...  Positions appear in the figure in absolute values.}%
\label{blue_fig}%
\end{figure}

Depending on the threshold values one obtains different dynamical patterns. When the values of  $\Gamma_1$ and $\Gamma_2$ are small only one or at most a few bottleneck passages take place before there is a change of periodic window. When these values are large the number of bottleneck passages is large before there is a change in periodic window, an indication that the system is robust to environmental variations. The dynamical properties of the model are sensitive to the imbalances represented by $\delta_1$ and $\delta_2$ and this sensitivity represents evolutionary changes. The repetition of this prescription leads to the dynamical behavior shown in Fig. \ref{blue_fig} that can be compared with that obtained from the TaNa model in Fig. \ref{fig:1}. The quasi-periodic episode of period $\tau_{n}$ is identified with the quasi stable co-existence of $n$ species for a time
period $T_{n}$ in the TaNa model and the chaotic burst at its ending leads
to some extinctions and new mutated species of the following quasi-stable configuration.

%This approach can be considered to be a phenomenological modelling of the
%original TaNa model. The threshold selections of the periodic windows $\tau_{n}$
%at $\nu _{\tau_{n}}$ and of the value of the control parameter distance $\delta
%\nu _{n}$ from the corresponding tangent bifurcation can be further
%elaborated, \textit{e.g.} by devising specific rules suggested by ecological
%principles associated with reproduction, mutation and death, and in this way
%obtain a closer reproduction of the dynamics of the TaNa model. Interestingly,
%an average decrement of  the variables $\delta \nu_{n}$ with increasing time $t$, that
%implies an average increment of the duration of quasi-stable episodes $T_{n}$
%with $t$, observed in the TaNa dynamical properties, signals an approach to
%the intermittency transition out of chaos. Our modelling by means of the
%dynamics associated with families of tangent bifurcations implies (in a
%well-defined manner restricted to deterministic nonlinear dynamics) that
%the ecological evolution model operates near the onset of chaos, in our case, at
%nearly vanishing Lyapunov exponent.

\section{Discussion and Conclusion}   
 We have considered two related high-dimensional model systems designed to represent evolving ecological systems where agents or strategies are species that undergo reproduction, death or mutation. One of them, the TaNa model is a fully stochastic model, whereas the other, a game-theoretic adaptation of the former, contains both deterministic and stochastic elements. Both models have been shown to display non-stationary intermittent behavior on macroscopic (many individual generation) time scales. Given that these model systems show macroscopic collective behavior reminiscent of low-dimensional nonlinear intermittency, we have attempted to extract from them expressions for simple nonlinear iterated maps by introducing approximations. The resulting low-dimensional dissipative maps display attractors associated with intermittency near tangent bifurcations.

The successive simplifications that have been introduced in modeling high-dimensi-onal complex systems and in exploring their properties follow this scheme: First, the TaNa model was built up by selecting simple mechanisms at the individual level for ecosystem evolution such as annihilation, reproduction and mutation to define basic time steps, and then time evolution lets these contribute to form more complicated interactions at the systems level. Second, the time evolution equations of a mean-field deterministic approximation of the TaNa model suggest a game-theoretic interpretation that leads to a replication-mutation model that preserves the non-stationary intermittent behavior for the macroscopic evolution, but permits considerations in the game theory language of strategies and pay-off values. Third, in discrete time space the replication-mutation model becomes a CML with stochastic terms so that the characteristics of the nonlinear maps that constitute it can be inspected. And finally, the latter problem was seen to represent a two-strategy symmetric game that within a time-discrete version constitutes a one-dimensional map with two control parameters. This was recognized \cite{DV_AR_AS_2011} as a replicator bimodal map that displays the routes to chaos familiar in unimodal maps that display period doublings, chaotic attractors and intermittency.

Therefore, pending stricter analysis, we preliminarily identify the macroscopic behavior of the high-dimensional model systems we describe here as composed of (effective) low-dimensional intermittency. The remarkable collapse of degrees of freedom that this circumstance entails may turn out to be more general than the few instances in which similar conduct has previously been encountered \cite{Chate_Manneville_1992,Ott_Antonsen_2008}. This prospect, and the advance in understanding it delivers, promotes a revitalization of a close relation between nonlinear dynamical theory and the science of complex systems.       
\bigskip

\bigskip AR and AD-R acknowledges support from DGAPA-UNAM-IN103814 and CONACyT-CB-2011-167978 (Mexican Agencies). HJJ was supported by the European project CONGAS (Grant FP7-ICT-2011-8-317672). DP and HJJ acknowledge interaction with Jelena Gruji{\'c}, at an early stage, concerning the replicator model.

\end{document}